\documentclass[sigconf]{acmart}
\usepackage{kotex}
\usepackage{placeins}
\usepackage{float}
\usepackage{microtype}
\usepackage{enumitem}
\usepackage{booktabs}
\usepackage{array}
\usepackage{multirow}
\usepackage{tabularx}
\usepackage{subcaption}
\usepackage{balance}
\usepackage{hyperref}
\usepackage{pgfplots}
\usetikzlibrary{patterns,patterns.meta}
\pgfplotsset{compat=1.18}
\usepackage[normalem]{ulem}
\usepackage{pgfplotstable}
\usepackage{mathtools}

\newcolumntype{C}{>{\centering\arraybackslash}p{2cm}}
\hypersetup{
    colorlinks=true,
    linkcolor=blue,
    urlcolor=cyan
}
\usepackage{soul}

\usepackage{tikz}
\usepackage{wrapfig}
\usepackage{capt-of}   
\usepgfplotslibrary{statistics} 


\definecolor{darkblue}{rgb}{0.0, 0.0, 0.55}
\definecolor{ao(english)}{rgb}{0.0, 0.5, 0.0}
\definecolor{coolblack}{rgb}{0.0, 0.18, 0.39}
\definecolor{purpleheart}{rgb}{0.41, 0.21, 0.61}
\definecolor{pastelviolet}{rgb}{0.8, 0.6, 0.79}
\definecolor{lightskyblue}{rgb}{0.53, 0.81, 0.98}
\definecolor{palecornflowerblue}{rgb}{0.67, 0.8, 0.94}
\definecolor{lightmauve}{rgb}{0.86, 0.82, 1.0}
\definecolor{lightpastelpurple}{rgb}{0.69, 0.61, 0.85}
\definecolor{bronze}{rgb}{0.8, 0.5, 0.2}
\definecolor{armygreen}{rgb}{0.29, 0.33, 0.13}
\definecolor{darkpowderblue}{rgb}{0.0, 0.2, 0.6}
\definecolor{falured}{rgb}{0.5, 0.09, 0.09}
\definecolor{outerspace}{rgb}{0.25, 0.29, 0.3}
\definecolor{tangerine}{rgb}{0.95, 0.52, 0.0}
\definecolor{seagreen}{rgb}{0.18, 0.55, 0.34}
\definecolor{springgreen}{rgb}{0.0, 1.0, 0.5}
\definecolor{applegreen}{rgb}{0.55,0.71,0.0}
\definecolor{amethyst}{rgb}{0.6,0.4,0.8}
\definecolor{amber}{rgb}{1.0,0.49,0.0}
\definecolor{darkgreen}{rgb}{0,0.4,0} 

\usepackage[table]{xcolor}
\usepackage{pgfplots}
\usetikzlibrary{plotmarks}
\usepgfplotslibrary{groupplots}

\AtBeginDocument{%
  }

\copyrightyear{2026}
\acmYear{2026}
\setcopyright{cc}
\setcctype{by}
\acmConference[CIKM '26]{Proceedings of the 35th ACM International Conference on Information and Knowledge Management}{November 07--11, 2026}{Rome, Italy}
\acmBooktitle{Proceedings of the 35th ACM International Conference on Information and Knowledge Management (CIKM '26), November 07--11, 2026, Rome, Italy}
\acmDOI{10.1145/3799682.3840824}
\acmISBN{979-8-4007-2539-5/2026/11}

\author{Donguk Park}
\orcid{0009-0007-5185-788X}
\affiliation{%
  \institution{Ulsan National Institute of Science and Technology (UNIST)}
  \city{Ulsan}
  \country{Republic of Korea}
}
\email{parkdw00@unist.ac.kr}

\author{Dongwon Lee}
\orcid{0000-0001-8371-7629}
\affiliation{%
  \institution{The Pennsylvania State University}
  \city{University Park}
  \state{PA}
  \country{USA}
}
\email{dongwon@psu.edu}

\author{Yeon-Chang Lee}
\orcid{0000-0002-8769-0678}
\authornote{Corresponding author.}
\affiliation{%
  \institution{Ulsan National Institute of Science and Technology (UNIST)}
  \city{Ulsan}
  \country{Republic of Korea}
}
\email{yeonchang@unist.ac.kr}
\begin{document}

\title{EchoTrace: Diagnosing Recursive Risks in LLM-Powered Recommender Systems}

\renewcommand{\shortauthors}{Donguk Park, Dongwon Lee, \& Yeon-Chang Lee}

\begin{abstract}
Large language models (LLMs) are increasingly integrated into recommender systems as data augmenters, profile generators, and recommendation modules. 
While these roles can enhance semantic understanding and recommendation quality, they also introduce LLM-specific risks such as bias and hallucination. 
These risks become more critical in feedback-loop settings, where LLM-generated signals and recommendations recursively shape future training data and recommendation dynamics.
In this paper, we propose a role-aware, phase-wise diagnostic framework for analyzing how LLM-induced risks emerge, propagate, and accumulate in LLM-powered recommender systems. 
Our framework combines controlled feedback-loop simulation with longitudinal phase-wise diagnosis across LLM-generated content, recommendation outputs, feedback-loop dynamics, and ecosystem-level effects.
Experiments on widely used benchmarks show that LLM-based components can amplify popularity bias, introduce spurious signals through hallucination, and gradually produce polarized and self-reinforcing exposure patterns over time. The code for EchoTrace
is available at \url{https://github.com/DongUk-Park/EchoTrace}.

\end{abstract}
%

%
\newcommand{\spec}{{\it spec.}}
\newcommand{\aka}{{\it a.k.a.}}
\newcommand{\ie}{{\it i.e.}}
\newcommand{\eg}{{\it e.g.}}
\newcommand{\llmrs}{\textsc{\textsf{LLM4RS}}}
\newcommand{\llmaug}{\texttt{LLM-as-Augmenter}}
\newcommand{\llmpro}{\texttt{LLM-as-Profiler}}
\newcommand{\llmrep}{\texttt{LLM-as-Representer}}
\newcommand{\llmrec}{\texttt{LLM-as-Recommender}}
\newcommand{\llmxai}{\texttt{LLM-as-XAI}}
\newcommand{\llmagent}{\texttt{LLM-as-RecAgent}}

\newcommand{\blue}{\textcolor{blue}}
\newcommand{\red}{\textcolor{red}}

\newcommand{\mj}[1]{\textcolor{blue}{[MJ: #1]}}
\newcommand{\jw}[1]{\textcolor{green}{[JW: #1]}}
\newcommand{\yc}[1]{\textcolor{red}{[YC: #1]}}


\newcommand{\oursall}{\textsc{\textsf{{TraceRec(all)}}}}
\newcommand{\oursforward}{\textsc{\textsf{{TraceRec(forward)}}}}
\newcommand{\oursrecent}{\textsc{\textsf{{TraceRec(recent)}}}}

\newcommand{\ourswoproj}{\textsc{\textsf{{TraceRec(w/o proj)}}}}

\maketitle
\section{Introduction} \label{sec:intro} \noindent\textbf{Background.} 
With the rapid growth of e-commerce and media platforms (\eg, Amazon and Netflix), users are exposed to an increasingly large and diverse content space, making recommender systems (RS) central to user experience and decision-making.
Despite the long-standing evolution of RS from collaborative filtering~\cite{cf_survey9, cf_content_survey10} to content-based and hybrid approaches~\cite{cf_content_survey10, hybrid_survey11}, these paradigms remain constrained by their reliance on \textit{limited observed signals}, such as user and item profiles and historical interaction data.


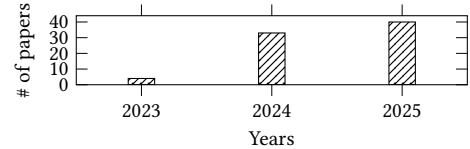
\begin{figure}[t]
\small
\begin{tikzpicture}
\begin{axis}[
    width=0.8\linewidth,
    height=2.5cm,
    ybar,
    bar width=10pt,
    ymin=0,
    enlarge x limits=0.25,
    symbolic x coords={2023,2024,2025, 2026$^\star$},
    xtick=data,
    axis line style={black},
    tick style={black},
    ylabel={\# of papers},
    xlabel={Years},
]
\addplot[
    fill=black!10,
    pattern=north east lines,
    pattern color=black
] coordinates {
    (2023,4)
    (2024,33)
    (2025,40)
    (2026$^\star$,39)
};
\end{axis}
\end{tikzpicture}
\vspace{-13pt}
\caption{Rapid growth of research on LLM-powered recommender systems in top-tier data mining and information retrieval conferences such as SIGIR, KDD, WWW, CIKM, and ICDM.
While this survey focuses on these venues, the overall research volume is substantially larger when accounting for additional conferences, journals, and preprints.
\\
\qquad $^\star$ Statistics for 2026 include publications through May.
}
\label{fig:pub}
\vspace{-0.35cm}
\end{figure}

To address these limitations, recent research has increasingly integrated \textit{large language models} (LLMs) into recommender systems~\cite{llm4rs_survey1_acmjournal, llm4rs_survey5_ieee24, llm4rs_survey6_kdd24}, giving rise to a new paradigm of \textbf{LLM-Powered RS} (\textbf{\llmrs}).
By leveraging extensive world knowledge and strong capabilities in text understanding and generation, LLMs enable recommender systems to infer richer semantic signals beyond explicitly observed interactions. 
This shift has been accompanied by a rapid surge of interest, as evidenced in Figure~\ref{fig:pub}.
Specifically, the number of studies grew from 4 in 2023, to 33 in 2024, and to 40 in 2025, with 39 additional studies already appearing in the first half of 2026, even when considering only top-tier venues in data mining and information retrieval.

\vspace{1mm}
\noindent\textbf{Taxonomy of LLM Roles.}
Within this emerging paradigm of \llmrs, LLMs are incorporated into recommendation pipelines through five functional roles: (R1) \llmaug, 
(R2) \texttt{LLM -as-Representer}, 
(R3) \texttt{LLM-as-Recommender}, (R4) \llmxai, and (R5) \llmagent.
We defer a detailed discussion of these roles to Section~\ref{sec:prelim} and briefly summarize them below.

\textbf{\llmaug} generates additional interaction signals to enrich training data, while \textbf{\texttt{LLM-as-Representer}} constructs representative user or item profiles from interaction histories and associated content.
\textbf{\texttt{LLM-as-Recommender}} directly generates or re-ranks recommendation outputs, while \textbf{\texttt{LLM-as-XAI}} provides natural language explanations for recommendations or model decisions.
Finally, \textbf{\llmagent} orchestrates agent-based pipelines that integrate multiple roles to perform end-to-end recommendation.

\vspace{1mm}
\noindent\textbf{Motivation.} 
However, this paradigm exhibits a paradoxical limitation: \textbf{its direct reliance on LLMs}.
Despite their strong capabilities, LLMs remain subject to a range of well-known issues, including \textit{hallucination}~\cite{FEF_hallucination_survey8_arxiv25} and \textit{bias}~\cite{Bias_survey14}.
Prior work in high-stakes domains such as medicine~\cite{medical_icml25} and law~\cite{legal_arxiv} has shown that these limitations can reinforce pre-existing beliefs and introduce false or misleading information into critical decision-making.
For instance, LLM-based legal systems have been reported to cite overruled or non-existent cases as valid precedents~\cite{legal_arxiv}.

These structural weaknesses can naturally \textit{translate to recommender systems}, where they pose risks to both recommendation reliability and algorithmic fairness.
For instance, in \texttt{LLM-as-Repr esenter}, incorrect inferences about users' sensitive attributes (\eg, gender or age) can introduce stereotype-driven recommendation signals that deviate from users' actual preferences.
Similarly, in \llmaug\ and \texttt{LLM-as-Recommender}, LLMs may overemphasize popularity signals or even generate hallucinated recommendations for non-existent items.

More critically, these risks do not remain isolated, but can accumulate through a \textbf{feedback loop}, in which LLM-influenced recommendation outputs are recursively incorporated as \textit{new training signals}.
Over time, such loops can reinforce LLM-induced biases and hallucinated signals within users' consumption patterns, gradually shifting the system from modeling human-driven behavior toward reinforcing LLM-induced signals. 
Consequently, this process may induce broader ecosystem-level effects, including \textit{polarization} that amplifies information and opinion gaps between groups ~\cite{polarization}. 
While feedback loops in traditional recommender systems have been studied extensively ~\cite{cf_feedback_loop1, cf_feedback_loop2}, those in \llmrs\ can introduce qualitatively different risks because LLM-induced signals themselves become part of future learning dynamics. 

\vspace{1mm}
\noindent\textbf{Our Work.} 
Therefore, we aim to \textit{empirically} investigate the LLM-induced risks of \llmrs\ and their long-term impacts on the recommendation ecosystem.
To this end, we first formulate a set of research questions centered on how LLM-induced risks evolve over repeated recommendation cycles.
Guided by these questions, we design a \textbf{controlled experimental pipeline} and a \textbf{three-phase diagnostic protocol} that together trace how LLM-induced signals propagate, accumulate, and transform over time.

Specifically, the \textbf{controlled experimental pipeline} formalizes a closed-loop setting in which LLM-based components influence both training and ranking processes, while recommendation outputs are recursively incorporated as future training signals.
To simulate recommendation dynamics where users consume a mixture of LLM-driven recommendations and human-driven interactions, the pipeline introduces a \textit{mixing coefficient} $\alpha$ that controls their relative proportion at each cycle. 
This design creates a controllable counterfactual environment in which such dynamics can be longitudinally observed and disentangled from organic user behavior.

Built on top of this pipeline, the \textbf{three-phase diagnostic protocol} performs targeted measurements across different stages of the recommendation lifecycle.
In the \textit{LLM Content Generation Phase}, we diagnose biases and hallucinations in LLM-generated content (LLMGC), such as user and item profiles and augmented interactions.
In the \textit{Recommendation Phase}, we evaluate popularity bias and hallucinated recommendations in LLM-based ranked lists.
In the \textit{Feedback Loop Phase}, we examine how these effects propagate across cycles, revealing long-term shifts in user behavior, item visibility, and polarization within the recommendation ecosystem.

\vspace{1mm}
\noindent\textbf{Contribution.} 
Our contributions establish a structured and extensible diagnostic framework for studying how LLM-induced signals emerge and propagate in \llmrs, as summarized below.
\begin{itemize}[leftmargin=*]
    \item \textbf{Functional Taxonomy:}
    We present a functional taxonomy of \llmrs\ based on a survey of $116$ recent studies from top-tier data mining and information retrieval venues, highlighting how LLMs intervene across the recommendation pipeline.
    \item \textbf{Problem Formulation:}
    We formulate the risks of \llmrs\ from a feedback-loop perspective, capturing both LLM-intrinsic and recursive user-model-data interaction dynamics.
    \item \textbf{Methodological Framework:}
    We propose a systematic framework consisting of a controlled experimental pipeline and a three-phase diagnostic protocol for longitudinal risk analysis.
    \item \textbf{Phase-Wise Empirical Study:}
    Using multiple \llmrs\ baselines and real-world datasets, we conduct extensive experiments to analyze how LLM-generated content exhibits bias and hallucination, how these effects manifest in recommendation outcomes, and how they propagate across recommendation cycles.
\end{itemize}
\section{Preliminaries} \label{sec:prelim} \begin{figure}[t]
    \centering
    \resizebox{\columnwidth}{!}{
        \usetikzlibrary{arrows.meta, positioning}

\definecolor{pBlue}{RGB}{240, 245, 255}
\definecolor{pGreen}{RGB}{240, 252, 240}
\definecolor{pYellow}{RGB}{255, 253, 235}
\definecolor{pOrange}{RGB}{255, 248, 240}
\definecolor{pPurple}{RGB}{250, 240, 255}
\definecolor{pGray}{RGB}{245, 245, 245}

\begin{tikzpicture}[
    node distance=0.4cm and 0.6cm,
    base/.style={draw, semithick, rounded corners=2pt, align=center, font=\rmfamily},
    role/.style={base, text width=4cm, minimum height=0.85cm, font=\large},
    ref/.style ={base, text width=6.7cm, minimum height=1cm, align=left, font=\normalsize},
    link/.style={draw, -{Stealth[length=1.3mm]}, thick, gray!35}
]

\node[role, fill=pBlue,   draw=blue!25]   (A) {\llmaug};
\node[ref,  fill=pBlue,   draw=blue!15,   right=of A] (RA) {\cite{aug_arxiv_1, aug_recsys24_2, aug_www25_3, augmentation_Wang_aug_www24_4, comb_aug_prof_cikm24_3, comb_aug_rep_wsdm25_16, comb_prof_aug_www25_17, comb_aug_xai_icassp24_21, comb_aug_prof_rep_wsdm24_20, comb_aug_prof_rep_recsys24_7, comb_aug_prof_rep_recsys24_9, llmrec_Wei_comb_aug_prof_rep_wsdm24_12, aug_10.1145/3746252.3761080}};
\draw[link] (A.east) -- (RA.west);

\node[role, fill=pGreen,  draw=green!25,  below=of A] (P)  {\llmrep};
\node[ref,  fill=pGreen,  draw=green!15,  right=of P] (RP) {\cite{prof_aaai25_3, prof_recsys24_1, comb_prof_rep_www24_13, prof_sigir24_2, prof_sigir25_4, comb_prof_rep_arxiv_1, comb_aug_prof_cikm24_3, comb_prof_aug_www25_17, comb_prof_rec_www24_24, comb_prof_rep_www24_23, comb_prof_rep_www24_27, comb_prof_rep_www24_25, comb_rec_prof_icde24_4, comb_rec_prof_www24_14, comb_prof_rep_cikm24_28, comb_prof_rep_recsys24_6, comb_prof_rep_recsys24_8, comb_prof_rep_sigir24_22, comb_prof_rep_sigir25_29, comb_prof_rep_sigir25_31, comb_prof_rep_sigir25_32, comb_prof_rec_icassp25_19, comb_prof_rep_coling25_18,  comb_aug_prof_rep_wsdm24_20, comb_aug_prof_rep_recsys24_7, comb_aug_prof_rep_recsys24_9, comb_prof_rec_xai_sigir25_30, comb_prof_rec_sigir25short_33, llmrec_Wei_comb_aug_prof_rep_wsdm24_12, rep_aaai25_6, rep_cikm24_1, rep_cikm24_8, rep_emnlp24_2, rep_neurips24_3, rep_recsys23_7,  rep_sigir25_10, rep_www24_9, comb_rec_rep_cikm24_2, comb_aug_rep_wsdm25_16, comb_rec_rep_recsys24_4, comb_rec_rep_sigir24_10, comb_prof_rep_sigir25_11, comb_rec_rep_xai_wsdm25_15,  rep_10.1145/3746252.3760952, rep_10.1145/3746252.3761284, rep_10.1145/3746252.3761369, rep_kdd25_2,rep_kdd25_3,rep_rec_10.1145/3746252.3760957,rep_rec_10.1145/3746252.3761214,rep_xai_10.1145/3746252.3761126, kang2026multi}};
\draw[link] (P.east) -- (RP.west);


\node[role, fill=pPurple, draw=purple!25, below=of P] (Rec) {\llmrec};
\node[ref,  fill=pPurple, draw=purple!15, right=of Rec] (RRec) {\cite{rec_acmrec25_9, rec_cikm23_1, rec_ecir24_2, rec_kdd24_3, rec_kdd24_4,rec_recsys23_5, rec_sigir24_6, rec_sigir25short_10, rec_wsdm25_7, rec_www25_8,comb_rec_rep_cikm24_2, comb_prof_rec_www24_24, comb_rec_prof_icde24_4, comb_rec_prof_www24_14, comb_rec_rep_sigir24_10, comb_rec_xai_sigir24_11, comb_prof_rec_icassp25_19, comb_rec_rep_xai_wsdm25_15, comb_prof_rec_xai_sigir25_30, comb_prof_rec_sigir25short_33, allmrec_Kim_comb_rec_rep_kdd24_5, rec_10.1145/3746252.3761280, rec_10.1145/3746252.3761384, rec_kdd25_1, comb_rec_rep_recsys24_4, contradiction_rec_emnlp25}};
\draw[link] (Rec.east) -- (RRec.west);

\node[role, fill=pOrange, draw=orange!25, below=of Rec] (E)  {\llmxai};
\node[ref,  fill=pOrange, draw=orange!15, right=of E] (RE) {\cite{comb_rec_xai_sigir24_11, comb_aug_xai_icassp24_21, comb_rec_rep_xai_wsdm25_15, comb_prof_rec_xai_sigir25_30, xai_10.1145/3746252.3761020, rep_xai_10.1145/3746252.3761126}};
\draw[link] (E.east) -- (RE.west);

\node[role, fill=pGray,   draw=gray!35, below=of E] (Agent) {\llmagent};
\node[ref,  fill=pGray,   draw=gray!25, right=of Agent] (RAgent) {\cite{multi_agent_www24_5, multi_agent_sigir24_1, macrec_multi_agent_sigir24_3, single_agent_sigir24_2, Zhang_single_agent_sigir24_4, multi_agent_sigir25short_6, single_agent_sigir25short_7, agent_10.1145/3746252.3760995, agent_10.1145/3746252.3761127}};
\draw[link] (Agent.east) -- (RAgent.west);

\end{tikzpicture}
    }
    \vspace{-15pt}
    \caption{Taxonomy of LLM roles in \llmrs, showing the distribution of all surveyed studies across role categories.}
    \label{fig:llm_taxonomy}
    \vspace{-0.35cm}
\end{figure}
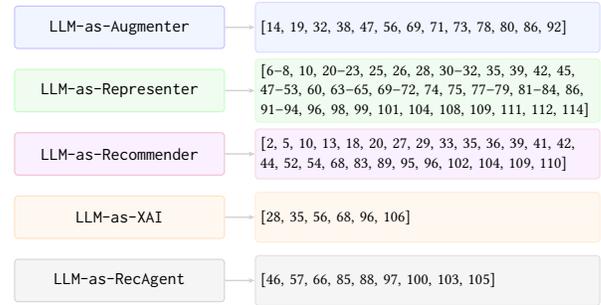
 
\subsection{LLM Roles in \llmrs}\label{sec:taxonomy} 
As shown in Figure~\ref{fig:pub}, recent years have witnessed rapid growth in \llmrs\ studies. 
While several survey works have reviewed \llmrs ~\cite{llm4rs_survey1_acmjournal, llm4rs_survey5_ieee24, llm4rs_survey6_kdd24}, they primarily focus on model architectures, datasets, and application scenarios, with limited attention to the \textit{functional roles} through which LLMs intervene across the recommendation pipeline.
To address this gap, we conduct a PRISMA-style systematic search ~\cite{prisma} using keywords ``\textit{LLM-powered recommender systems},'' ``\textit{LLM for recommendation},'' and ``\textit{LLM-based recommender system}'' on Google Scholar and Semantic Scholar, identifying 116 \llmrs\ studies from top-tier venues (2023--2026.05).
Based on this analysis, we identify five functional \textit{roles}, as illustrated in Figure~\ref{fig:llm_taxonomy}, and organize them into two high-level functional \textit{objectives}: content-generation and decision-making.

First, content-generation roles produce intermediate signals used for model training as follows:

\begin{itemize}[leftmargin=*]
    \item \textbf{\llmaug:}
    This role generates additional user-item interaction signals that are not explicitly available in the original dataset, thereby enriching the training data.
    For example, Wang et al.~\cite{augmentation_Wang_aug_www24_4} prompt an LLM with a user's interaction history and a randomly sampled pair of cold items (\ie, items with zero interactions in the training data), asking the model to select the item that better matches the user's latent preferences.
    The selected items are then incorporated as synthetic interactions.

    \item \textbf{\llmrep:}
    This role constructs representative user or item abstractions, including high-level profiles and embedding representations, by leveraging interaction histories and associated content.
    For example, Wei et al.~\cite{llmrec_Wei_comb_aug_prof_rep_wsdm24_12} generate user profiles capturing explicit preferences (\eg, liked and disliked genres) from historical interactions and item attributes, which are then transformed into user embeddings.
\end{itemize}

Next, decision-making roles directly influence recommendation outputs and user exposure as follows:

\begin{itemize}[leftmargin=*]
    \item \textbf{\llmrec:}
    This role directly produces recommendation outputs by re-ranking a candidate set retrieved by a backbone RS, or by generating item identifiers or titles without an explicit candidate pool.
    For instance, Bao et al.~\cite{rec_acmrec25_9} operate in a candidate-free setting, prompting the LLM to generate item titles directly from interaction histories.

    \item \textbf{\llmxai:} 
    This role provides natural language explanations for either recommendation outputs or intermediate LLMGC, aiming to enhance transparency and user trust.
    For example, Ren et al.~\cite{comb_prof_rep_www24_13} produce rationales underlying user and item profile construction, thereby supporting both interpretability and recommendation performance.

    \item \textbf{\llmagent:}
    This role introduces LLM-based agents that autonomously orchestrate the recommendation pipeline, either through a single unified agent or multiple specialized agents that collaboratively solve sub-tasks.
    For example, Wang et al.~\cite{macrec_multi_agent_sigir24_3} design a collaborative framework where agents specialize in user analysis, item analysis, retrieval, and decision management, and coordinate to perform recommendation.

\end{itemize}

\begin{figure*}[t]
\centering
\includegraphics[width=0.96\linewidth]{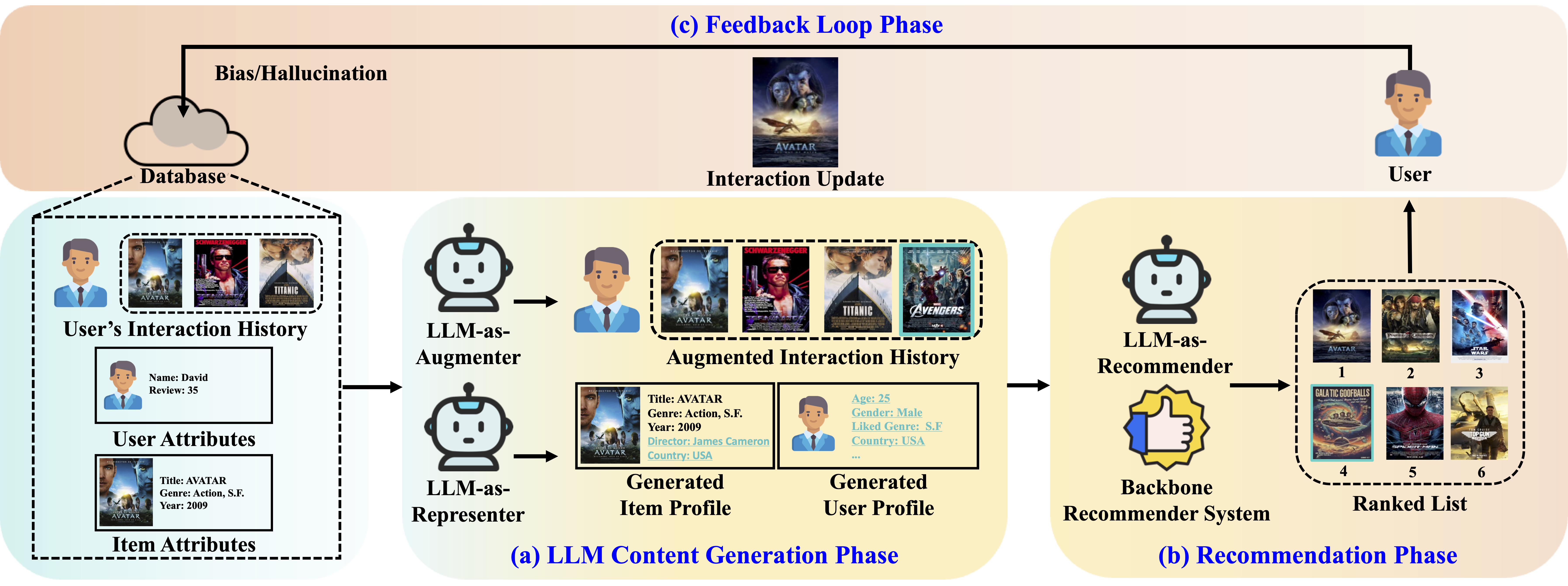}
\vspace{-0.4cm}
\caption{Overview of the proposed diagnostic framework under a controlled feedback-loop pipeline. The framework consists of three phases: (P1) LLM Content Generation, (P2) Recommendation, and (P3) Feedback Loop.
} \label{fig:overview}
\vspace{-0.35cm}
\end{figure*}

The roles defined above are often combined in a \textbf{compositional manner}. 
Multiple roles may operate within a single prompt, or LLMGC from one role may serve as input to another, forming multi-stage recommendation pipelines.
For instance, Ren et al.~\cite{comb_prof_rep_www24_13} combine \llmrep\ and \llmxai\ to generate user profiles with both preference summaries and rationales.
As a result, LLM-induced signals generated at one stage can influence downstream recommendation processes through subsequent roles.

\subsection{Risk Hypotheses in \llmrs}\label{sec:RiskScenario}
LLMs exhibit well-known limitations, including bias and hallucination~\cite{Bias_survey14, FEF_hallucination_survey8_arxiv25}, that can propagate beyond textual outputs and affect downstream systems across diverse domains~\cite{medical_icml25, legal_arxiv}.

In \llmrs, similar risk transfer mechanisms can arise when LLMs are embedded into the recommendation pipeline through different functional roles, where their effects can manifest differently across pipeline stages.
Accordingly, we structure our \textbf{risk hypotheses} into three phases: \textit{content-generation}, \textit{decision-making}, and \textit{feedback-loop}.
Under this formulation, we focus on three widely adopted roles: \llmaug, \llmrep, and \texttt{LLM -as-Recommender}, as they directly influence either training signals or recommendation outputs and represent dominant design patterns in the \llmrs\ literature (see Figure~\ref{fig:llm_taxonomy}).

\vspace{1mm}
\noindent\textbf{H1: Risks in Content Generation.} 
Content-generation roles, such as \llmaug\ and \texttt{LLM-as-Representer}, can \textit{distort the training signals} by amplifying existing biases or introducing unsupported LLM-generated attributes and preferences.
For example, LLM-generated profiles or augmented interactions may assign strong preferences or attributes even when the interaction history provides limited supporting evidence.
In more severe cases, hallucinated attributes (\ie, unsupported user occupations or item genres) may induce \textit{artificial user-item associations} absent from the original data.
Accordingly, we hypothesize that these effects can be observed within the \textit{LLMGC itself} as skewed preference patterns and hallucination-induced user-item associations.

\vspace{1mm}
\noindent\textbf{H2: Risks in Decision Making.} 
Decision-making roles, such as \texttt{LLM-as-Recommender}, can \textit{produce unreliable recommendation outputs} when LLMs directly select or re-rank items. 
For example, they may generate invalid or non-existent items, exhibit inconsistent ranking behavior under identical inputs, or disproportionately favor more popular items.
Accordingly, we hypothesize that these effects can be observed in the \textit{final ranked lists} as hallucinated recommendations, ranking instability, and skewed item exposure.

\vspace{1mm}
\noindent\textbf{H3: Risks under Feedback Loop.}
When LLM-generated content and recommendation outputs are repeatedly consumed by users and incorporated back into training, they form a \textit{feedback loop} in which the risks in H1 and H2 can accumulate and evolve over time. 
As recommendation cycles repeat, \llmrs\ may gradually shift from modeling human-driven behavior toward reinforcing LLM-induced signals.
This dynamic can amplify existing biases, introduce artificial behavioral patterns, and reshape exposure across users and items.
Accordingly, we hypothesize that these effects can manifest as increasing \textit{group separation} in user or item embeddings and \textit{skewed exposure} across user or item groups.

In summary, the risks of \llmrs\ can arise not only from LLM limitations, but also from how LLM-induced signals propagate through repeated recommendation cycles.
Building on these hypotheses, we formulate the following research questions (RQ):
\begin{itemize}[leftmargin=*]
    \item \textbf{RQ1 (LLMGC-Level):}  
    How does LLM-generated content exhibit bias or hallucination when used as training inputs in \llmrs?
    \item \textbf{RQ2 (Ranking-Level):}  
    How do LLM-based decision-making roles affect the reliability and stability of recommendation outputs, as reflected in item validity and ranking consistency?
    \item \textbf{RQ3 (Feedback-Loop-Level):}  
    As LLM-generated content and recommendations accumulate over time, how do the effects identified in RQ1 and RQ2 propagate through feedback loops?
    \item \textbf{RQ4 (Ecosystem-Level):}  
    What broader, long-term risks do these dynamics pose to the recommendation ecosystem?
\end{itemize}

However, prior work~\cite{llm4rs_survey1_acmjournal, llm4rs_survey6_kdd24, llm4rs_survey5_ieee24} lacks a systematic framework for examining how these risks emerge across functional roles and evolve through repeated recommendation cycles under controlled settings.
Accordingly, our goal is to establish a structured, longitudinal framework that enables these dynamics to be \textit{empirically observed, analyzed, and tracked over time}.

\section{The Proposed Framework: \texttt{EchoTrace}} \label{sec:method} In this section, we present \texttt{EchoTrace}, a framework for examining how LLM-induced risks in \llmrs\ emerge, propagate, and accumulate over repeated recommendation cycles. 
To this end, we introduce a controlled feedback-loop pipeline (\textsection\ref{sec:FeedbackLoop-Pipeline}) and three-phase diagnostic protocol (\textsection\ref{sec:diagnostics}) for systematic longitudinal risk analysis.
 
\subsection{Overview}\label{sec:method-overview} 
We provide a high-level overview of our diagnostic framework, 
which follows the lifecycle of an \llmrs\ through three successive phases (Figure~\ref{fig:overview}).
The process begins with the \textbf{LLM Content Generation Phase} (Figure~\ref{fig:overview}-(a)), where content-generation roles (\ie, \texttt{LLM-as-Augmenter}\ and \llmrep) transform users' historical interactions into intermediate signals, such as user and item profiles or augmented interactions.
When partial user or item attributes are available, LLMs may also infer missing attributes to complete the profiles.
At this phase, our diagnosis examines whether LLMGC exhibits bias, hallucination, or systematic distortions that deviate from the underlying interaction data.

Next, in the \textbf{Recommendation Phase} (Figure~\ref{fig:overview}-(b)), decision-making roles (\ie, \llmrec) determine the final recommendation outputs presented to users, either by re-ranking candidates retrieved by a backbone RS (\eg, LightGCN~\cite{lightgcn}) or by generating recommendations in an open-ended manner.
In \llmrs\ that employ both LLMGC and decision-making roles, the generated content may also be incorporated into model training prior to recommendation.
At this phase, our diagnosis focuses on how LLM-induced decisions manifest in user-facing outcomes, including unreliable rankings, hallucinated recommendations, and skewed exposure patterns such as popularity bias.

Finally, the \textbf{Feedback Loop Phase} (Figure~\ref{fig:overview}-(c)) reintegrates user interactions influenced by LLMGC and LLM-driven rankings as training signals for subsequent cycles.
At this phase, our diagnosis examines how distortions from earlier phases accumulate over time, potentially shifting the system from modeling human-driven preferences toward reinforcing LLM-induced patterns.

\subsection{Feedback Loop Pipeline}\label{sec:FeedbackLoop-Pipeline} 
\begin{figure}[t] 
\centering
\includegraphics[width=0.95\linewidth]{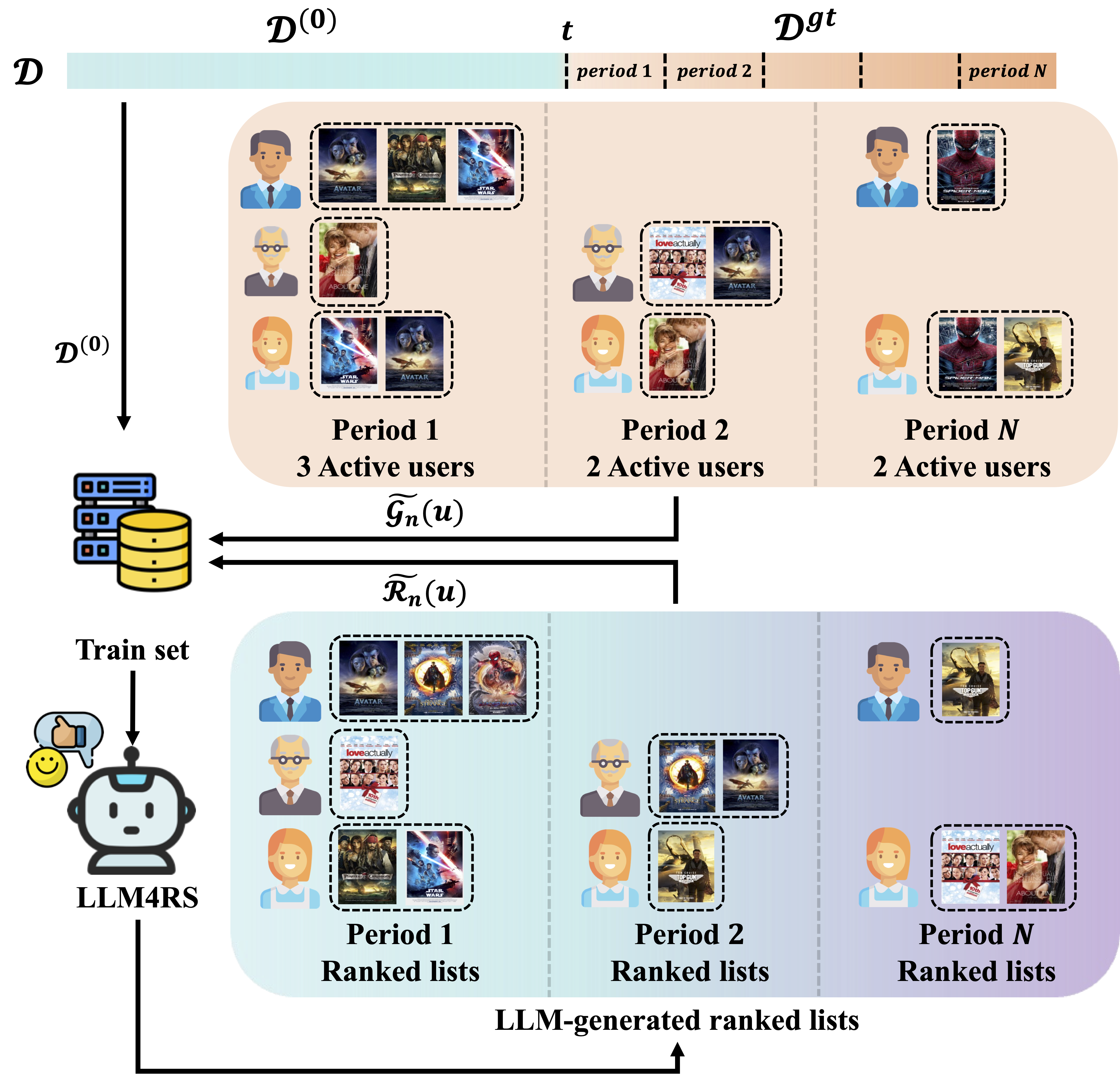}
\vspace{-0.4cm}
\caption{Overview of the controlled feedback-loop simulation pipeline. 
The timeline after the cutoff time $t$ is partitioned into $N$ consecutive periods based on the ground-truth interaction log.
In each period, recommendations are generated only for active users identified from the ground truth.}
\label{fig:FeedbackLoop}
\vspace{-0.4cm}
\end{figure}

In this subsection, we introduce a controlled feedback-loop pipeline that simulates how LLM-driven recommendations recursively reshape future interaction data over time. 
As illustrated in Figure~\ref{fig:FeedbackLoop}, this pipeline constructs a closed-loop process in which recommendation outputs are continuously reincorporated into training through repeated recommendation cycles.

\vspace{1mm}
\noindent\textbf{Temporal Split \& Initialization.}
Let $\mathcal{U}$ and $\mathcal{I}$ denote the sets of users and items, respectively. 
We define the full interaction log as $\mathcal{D}=\{(u, i, \tau) \mid u \in \mathcal{U},\; i \in \mathcal{I},\; \tau \in \mathbb{R}^{+}\}$, where each tuple $(u, i, \tau)$ denotes an interaction between user $u$ and item $i$ at timestamp $\tau$.

Given a temporal cutoff $t$, we partition $\mathcal{D}$ into an initial training set and a ground-truth activity set (\ie, test set) as follows:
\begin{equation}\label{eq:partition}
    \begin{aligned}
        \mathcal{D}^{(0)} = \{(u, i, \tau) \in \mathcal{D} \mid \tau \leq t\}, \quad
        \mathcal{D}^{\text{gt}} = \{(u, i, \tau) \in \mathcal{D} \mid \tau > t\},
    \end{aligned}
\end{equation}
where $\mathcal{D}^{(0)}$ initializes the \llmrs, while $\mathcal{D}^{\text{gt}}$ is used to preserve the distribution of user activity (\eg, interaction timing and frequency) during simulation.
We also define the set of \emph{common users} as $\mathcal{U}_{\text{com}} = \{u\mid\exists (u, i, \tau) \in \mathcal{D}^{(0)} \;\land\; \exists (u, j, \tau') \in \mathcal{D}^{\text{gt}}\}$, \ie, users appearing in both sets.

\vspace{1mm}
\noindent\textbf{Recursive Recommendation \& Injection.}
The post-$t$ timeline is divided into $N$ consecutive periods of equal duration.
Let $\mathcal{D}^{(n)}$ denote the accumulated interaction data in period $n$, where $\mathcal{D}^{(0)}$ is the initial training set defined in Eq.~(\ref{eq:partition}).
For each period $n \in \{1, \cdots,~N\}$, the pipeline proceeds as follows:

\begin{enumerate}[leftmargin=*, itemsep=2pt]
    \item \textbf{Recommend:}  
    During period $n$, we define the set of active users $\mathcal{U}_n$ as the subset of \emph{common users} who exhibit at least one interaction in $\mathcal{D}^{\text{gt}}$ \textit{within} period $n$.
    For each active user $u \in \mathcal{U}_n$, the system produces a Top-$K_u$ recommendation list $\mathcal{R}_n(u)$ in period $n$, where $K_u$ matches the number of items that $u$ actually consumed \textit{during} period $n$ in $\mathcal{D}^{\text{gt}}$.
    This preserves each user's activity volume within the period.

    \item \textbf{Inject:} 
    By default, recommended items are treated as observed interactions and appended to the dataset, yielding $\mathcal{D}^{(n)} = \mathcal{D}^{(n-1)} \cup \{(u, i, \tau_n) \mid u \in \mathcal{U}_n,\; i \in \mathcal{R}_n(u)\},$ where $\tau_n$ is the canonical timestamp for period $n$.
    To additionally simulate mixed-consumption settings, we introduce a mixing coefficient $\alpha \in [0,1]$. 
    Instead of injecting the full recommendation list, only a fraction $\alpha$ of recommended items is incorporated into the feedback loop, while the remaining interactions are randomly sampled from the user's ground-truth activity in the corresponding period.
    Formally, the injected interaction set is updated as follows:   
    \begin{equation}\label{eq:alpha-inject}
    \mathcal{D}^{(n)} = \mathcal{D}^{(n-1)} \cup \{ (u, i, \tau_n) \mid u \in \mathcal{U}_n,\; i \in \widetilde{\mathcal{R}}_n(u) \cup \widetilde{\mathcal{G}}_n(u) \},
    \end{equation}
    where $\widetilde{\mathcal{R}}_n(u)$ and $\widetilde{\mathcal{G}}_n(u)$ denote the sampled recommendation and ground-truth subsets, respectively, with $|\widetilde{\mathcal{R}}_n(u)| = \lfloor \alpha K_u \rfloor$ and $|\widetilde{\mathcal{G}}_n(u)| = K_u - \lfloor \alpha K_u \rfloor$. 
    Thus, $\alpha = 1$ corresponds to the default fully LLM-driven setting, while smaller values simulate mixed-consumption settings; 
    for example, $\alpha = 0.5$ indicates that half of the injected interactions originate from LLM-driven recommendations and the other half from human-driven interactions.
    This formulation primarily supports controlled analysis under the fully LLM-driven setting ($\alpha = 1$), while also enabling additional investigation of mixed-consumption scenarios with partial human-driven interactions.

    \item \textbf{Train:}  
    After period $n$ concludes (\ie, at the beginning of period $n+1$), \llmrs\ is re-trained on $\mathcal{D}^{(n)}$, which consists of 
    $\mathcal{D}^{(0)}$ and all injected interactions accumulated up to period $n$.
    Here, even though only interaction data is explicitly injected into $\mathcal{D}^{(n)}$, the effects of LLMGC generated in the content-generation phase can still propagate through retraining, as LLM-derived profiles and augmented signals influence the learned representations used for subsequent recommendations.
\end{enumerate}

By iterating this process across all periods, the pipeline constructs a controlled longitudinal simulation that preserves the original temporal activity volume of users while progressively incorporating \llmrs-induced interactions into the feedback loop.
This setting enables the following counterfactual question: 
\vspace{1mm}
\begin{center}
\fbox{
\parbox{0.95\linewidth}{
\centering
\textit{How would \llmrs\ evolve as increasing portions of user interactions become driven by LLM-based recommendations and recursively incorporated into training?}
}
}
\end{center}

\subsection{Phase-Wise Risk Diagnostics}\label{sec:diagnostics} 
In this subsection, we introduce a phase-wise diagnostic protocol that examines how LLM-induced risks emerge, propagate, and accumulate throughout the lifecycle of \llmrs. 
The protocol is organized into three successive phases corresponding to different stages of the recommendation process, thereby explicitly aligning our analysis with the four levels of research questions, \ie, LLMGC, ranking, feedback-loop, and ecosystem levels.
This design enables risks to be diagnosed under controlled conditions at each stage while also revealing how early-stage distortions can later manifest as long-term system-level effects.

Across all phases, we focus on two primary intrinsic risk dimensions, \ie, \textit{bias} and \textit{hallucination}, because they are consistently observable across diverse LLM roles, directly transferable from content generation to recommendation outcomes, and repeatedly reinforced through feedback loops.
Moreover, both risks can be quantitatively measured and longitudinally tracked across recommendation cycles, making them particularly suitable for controlled diagnosis.
In contrast, other LLM risks, such as \textit{prompt sensitivity}~\cite{promptsensitivity_ACL25} or \textit{sycophancy}\footnote{Sycophancy refers to the tendency of a model to overly align with the user's intent or input at the expense of objective correctness.}~\cite{sycophancy_aaai25}, typically require adversarial prompting environments, conversational user settings, or external attack protocols, which are beyond the scope of this work.

\begin{itemize}[leftmargin=*]
    \item \textbf{Bias:} 
    We define bias as systematic imbalances in the distributions of LLMGC or recommendation outputs. 
    These imbalances may arise from both pre-existing behavioral patterns and LLM-induced effects, such as amplified popularity bias or skewed user/item attribute distributions absent from the original data.
    \item \textbf{Hallucination:} 
    We define hallucination as the generation of information that is not supported by the observed interaction data or the system's item pool. 
    This includes \textit{factual errors and fabrications (FEF)}~\cite{FEF_hallucination_survey8_arxiv25}, such as non-existent user attributes or items, as well as \textit{logical contradictions (LC)}~\cite{contradiction_rec_emnlp25}, where repeated executions under identical conditions produce inconsistent outputs.
\end{itemize}

Detailed definitions of all evaluation metrics are provided in Section~\ref{sec:setting}.
Although this study focuses on \textit{bias} and \textit{hallucination}, the proposed framework is modular and extensible.
Since each phase operates as an independent diagnostic component, additional risk dimensions can be incorporated through new metrics without modifying the overall pipeline.
Accordingly, the framework can serve as a \textbf{general diagnostic protocol} for diverse \llmrs.

\subsubsection{Phase 1 Diagnosis (LLMGC-Level, RQ1)}
This phase corresponds to the stage where LLMs operate in content-generation roles and produce intermediate signals such as user/item profiles or augmented interactions.
The goal is to examine whether LLMGC already contains intrinsic distortions that may later propagate into recommendation outputs and feedback-loop dynamics.

\vspace{1mm}
\noindent\textbf{Bias in LLMGC.}
We analyze the distributions of generated user/item profiles and augmented interactions, and compare them with the original data distributions. 
Specifically, we examine whether LLMGC (i) amplifies existing skews, such as dominant genres or user attributes, or (ii) introduces new distributional patterns absent from the ground-truth observations.
This analysis enables early-stage diagnosis of bias transfer before recommendation generation. 

\vspace{1mm}
\noindent\textbf{Hallucination in LLMGC.}
We diagnose hallucination using two complementary metrics. 
First, FEF rate quantifies generated attributes or interactions that do not exist in the dataset, such as non-existent occupations or invalid candidate items.
Second, LC rate measures inconsistencies across repeated executions under identical inputs, such as conflicting inferred profile attributes or unstable candidate-item selections.
This analysis enables diagnosis of hallucinated signals before the ranking stage.

\subsubsection{Phase 2 Diagnosis (Ranking-Level, RQ2)}
This phase corresponds to the stage where recommendation outputs are produced either directly by decision-making roles or indirectly by models trained on LLMGC. 
The goal is to examine how the intrinsic risks identified in Phase 1 manifest in the reliability and exposure patterns of ranked recommendation lists.

\vspace{1mm}
\noindent\textbf{Bias in Ranking.}
We quantify whether recommendation outputs become disproportionately concentrated on popular items relative to ground-truth interactions. 
For each user, we compare the average popularity of recommended items with that of actually consumed items.
The resulting popularity gap indicates the extent to which the ranking process shifts exposure toward dominant items.

\vspace{1mm}
\noindent\textbf{Hallucination in Ranking.}
We examine whether ranked lists contain hallucinated outputs.
Specifically, FEF captures invalid or non-existent recommendations, such as fabricated item titles generated by \texttt{LLM-as-Recommender}, while LC measures inconsistencies in ranked outputs across repeated executions under identical conditions.
This analysis enables diagnosis of potential trade-offs between exposure diversity and recommendation reliability in open-ended recommendation settings.

\subsubsection{Phase 3 Diagnosis (Feedback-Loop \& Ecosystem-Levels, RQ3-RQ4)}
This phase corresponds to the stage where recommendation outputs are recursively reintegrated through the feedback-loop pipeline.
The goal is to examine whether the risks identified in earlier phases persist, accumulate, and eventually induce long-term structural changes in the recommendation ecosystem.

\vspace{1mm}
\noindent\textbf{Accumulation of Bias and Hallucination.}
We track whether popularity-oriented distortions and hallucinated signals intensify across recommendation cycles.
Specifically, we measure temporal changes in popularity gaps together with longitudinal trends in FEF and LC rates.
This analysis enables diagnosis of whether early-stage risks become amplified through recursive feedback.

\vspace{1mm}
\noindent\textbf{Long-Term Structural Separation and Self-Reinforcement.}
We further diagnose whether repeated feedback-loop interactions induce structural polarization and self-reinforcing dynamics in the recommendation ecosystem.
To this end, we analyze the evolution of user and item embeddings across recommendation cycles.
Specifically, we examine whether different user or item groups become increasingly separated in the learned representation space by visualizing embedding structures and measuring changes in Euclidean distances between group centroids over time.
We further investigate whether the resulting separation aligns with conventional factors such as user preference divergence or popularity-based exposure, or instead reflects increasing reliance on previously injected LLM-induced signals.
This analysis enables diagnosis of whether repeated exposure to LLM-driven signals gradually produces increasingly separated and self-reinforcing ecosystem structures.

\section{Evaluation} \label{sec:eval} \subsection{Experimental Settings}\label{sec:setting} 
\vspace{1mm}
\noindent\textbf{Datasets.}
We evaluate the proposed framework on three widely used \llmrs\ benchmarks: MovieLens-1M (ML-1M), Amazon-Books (A-Books), and Yelp.
ML-1M provides user-item interaction logs along with explicit user (\eg, age, gender, occupation) and item (\eg, genres, titles) attributes. 
A-Books and Yelp also provide interaction logs and item attributes (\eg, titles, brands, categories for A-Books; business names, categories, locations for Yelp), but include limited explicit user attributes.
Table~\ref{Table:dataset} summarizes dataset statistics.



\begin{table}[t]
\centering
\caption{Dataset statistics}
\vspace{-0.2cm}
\label{Table:dataset}
\renewcommand{\arraystretch}{1.2}
\resizebox{\columnwidth}{!}{%
\begin{tabular}{lrrrrr}
\toprule
\textbf{Dataset} & \textbf{\# Users} & \textbf{\# Items} & \textbf{\begin{tabular}[c]{@{}r@{}}\# Train\\ Interactions\end{tabular}} & \textbf{\begin{tabular}[c]{@{}r@{}}\# Ground-Truth\\ Interactions\end{tabular}} & \textbf{Time Period} \\ \midrule
\textbf{ML-1M} & 5,400  & 3,693  & 800,167 & 200,042 & 2000.04.26. - 2003.03.01. \\ 
\textbf{A-Books} & 34,375 & 25,882 & 473,266 & 473,265 & 2017.01.01. - 2018.01.01. \\
\textbf{Yelp} & 12,025 & 11,233 & 80,217 & 52,027 & 2018.01.01. - 2018.12.31. \\ \midrule
\end{tabular}%
}
\vspace{-0.5cm}
\end{table}
\begin{table}[t]
\small
\centering
\caption{LLM backbones and estimated cost per feedback-loop period for \llmrs\ roles. For each role, we adopt the LLM backbone specified in the corresponding baseline implementation. The reported zero cost for \llmrec\ reflects the use of a free, open-source LLM backbone.}
\vspace{-0.2cm}
\label{Table:llm_statistics}
\renewcommand{\arraystretch}{1.3} 
\resizebox{\columnwidth}{!}{%
\begin{tabular}{cccc} 
\toprule
\textbf{LLM Role} & \textbf{LLM Backbone} & \textbf{Cost per Period} \\ \midrule
\llmaug & \texttt{gpt-4o} & \$15.6 (ML-1M) / \$27.45 (A-Books) / \$14.2 (Yelp) \\ 
\addlinespace[0.1cm] 
\llmrep & \texttt{gpt-4o} & \$13.8 (ML-1M) / \$122.4 (A-Books) / \$46.6 (Yelp) \\
\addlinespace[0.1cm]
\llmrec & \texttt{facebook/opt-6.7b} & \$0 (ML-1M) / \$0 (A-Books) / \$0 (Yelp) \\
\midrule
\end{tabular}%
}
\vspace{-0.35cm}
\end{table}
\vspace{1mm}
\noindent\textbf{\llmrs\ Baselines.}
We select representative \llmrs\ baselines corresponding to three widely adopted LLM roles:
\begin{itemize}[leftmargin=*]
    \item \textbf{Cold-Item Aug~\cite{augmentation_Wang_aug_www24_4}
    (\ie, \texttt{LLM-as-Augmenter}):}
    This method generates synthetic interactions for cold-start items and injects them into the training data of a conventional RS, representing \textit{data augmentation} roles.
    
    \item \textbf{LLMRec ~\cite{llmrec_Wei_comb_aug_prof_rep_wsdm24_12} (\ie, \llmrep):} 
    This method generates user and item profiles from interaction histories and item attributes, representing intermediate \textit{representation} roles in the recommendation pipeline.

    \item \textbf{A-LLMRec ~\cite{allmrec_Kim_comb_rec_rep_kdd24_5} (\ie, \llmrec):}
    This method directly generates ranked recommendation lists using an LLM, representing \textit{decision-making} roles at the recommendation stage.
\end{itemize}

\begin{figure}[t] 
\small
    \centering
    \resizebox{\columnwidth}{!}{ 
        \begin{tikzpicture}
            \begin{groupplot}[
                group style={
                    group size=3 by 3,       
                    horizontal sep=0.45cm,    
                    vertical sep=1.25cm,      
                },
                ybar,
                width=0.42\columnwidth,      
                height=2.3cm,                
                ylabel style={font=\tiny},
                xtick=data,
                xticklabel style={
                    rotate=45,
                    anchor=north east,
                    font=\fontsize{4}{5}\selectfont,
                    inner sep=1pt
                },
                yticklabel style={font=\fontsize{4.5}{5.5}},
                ymin=0,
                enlarge x limits=0.15,       
                ymajorgrids=true,
                grid style=dashed,
                title style={font=\fontsize{5.3}{6.3}\selectfont, yshift=-2pt},
            ]


            \nextgroupplot[title={\textbf{Gender (ML-1M)}}, ymax=6500,
                symbolic x coords={Male, Female}, enlarge x limits=0.5, bar width=10pt]
            \addplot[fill=seagreen!60!white, draw=black] 
                coordinates {(Male, 4641) (Female, 759)};

            \nextgroupplot[title={\textbf{Liked Genre (ML-1M)}}, 
                bar width=5pt, symbolic x coords={Drama, Comedy, Action, Sci-Fi, Thriller, Horror}, ymax=2400]
            \addplot[fill=seagreen!60!white, draw=black] coordinates {(Drama, 1949) (Comedy, 1224) (Action, 865) (Sci-Fi, 589) (Thriller, 222) (Horror, 138)};

            \nextgroupplot[title={\textbf{Liked Director (ML-1M)}}, 
                bar width=5pt, symbolic x coords={S. Spielberg, R. Scott, J. Cameron, Q. Tarantino, F. F. Coppola, W. Allen}, ymax=2300]
            \addplot[fill=seagreen!60!white, draw=black] coordinates {(S. Spielberg, 1889) (R. Scott, 788) (J. Cameron, 680) (Q. Tarantino, 412) (F. F. Coppola, 381) (W. Allen, 338)};


            \nextgroupplot[title={\textbf{Gender (A-Books)}}, ymax=38000, 
                symbolic x coords={Male, Female}, enlarge x limits=0.5, bar width=10pt]
            \addplot[fill=lightskyblue, draw=black] 
                coordinates {(Male, 4035) (Female, 28524)};

            \nextgroupplot[title={\textbf{Liked Category (A-Books)}}, 
                bar width=5pt, symbolic x coords={Romance, Mystery, Thriller, Literature, Sci-Fi, Bibles}, ymax=14500]
            \addplot[fill=lightskyblue, draw=black] coordinates {(Romance, 12382) (Mystery, 6471) (Thriller, 5433) (Literature, 4170) (Sci-Fi, 1304) (Bibles, 1128)};

            \nextgroupplot[title={\textbf{Liked Author (A-Books)}}, 
                bar width=5pt, symbolic x coords={J. Patterson, L. J. Shen, T. S. Joyce, M. Sullivan, M. March, B. Forrest}, ymax=320]
            \addplot[fill=lightskyblue, draw=black] coordinates {(J. Patterson, 261) (L. J. Shen, 214) (T. S. Joyce, 207) (M. Sullivan, 190) (M. March, 179) (B. Forrest, 165)};

        
            \nextgroupplot[
                title={\textbf{{Gender (Yelp)}}}, ymax=9000, 
                symbolic x coords={Male, Female}, enlarge x limits=0.5, bar width=10pt,
                scaled y ticks=false, yticklabel style={/pgf/number format/fixed}
            ]
            \addplot[fill=orange!40!white, draw=black] 
                coordinates {(Male, 7662) (Female, 4358)};

            \nextgroupplot[
                title={\textbf{Liked Category (Yelp)}}, bar width=5pt, 
                symbolic x coords={Restaurants, {American(New)}, Bars, Nightlife, Seafood, {Cajun/Creole}}, 
                ymax=4000,
                scaled y ticks=false, yticklabel style={/pgf/number format/fixed}
            ]
            \addplot[fill=orange!40!white, draw=black] 
                coordinates {(Restaurants, 3567) ({American(New)}, 711) (Bars, 692) (Nightlife, 656) (Seafood, 563) ({Cajun/Creole}, 554)};

            \nextgroupplot[
                title={\textbf{Liked City (Yelp)}}, bar width=5pt, 
                symbolic x coords={{Florida(Fl)}, {Pennsylvania(Pa)}, Philadelphia, {Louisiana(La)}, {New Orleans}, {Indiana(In)}}, 
                ymax=320,
                scaled y ticks=false, yticklabel style={/pgf/number format/fixed}
            ]
            \addplot[fill=orange!40!white, draw=black] 
                coordinates {({Florida(Fl)}, 261) ({Pennsylvania(Pa)}, 214) (Philadelphia, 207) ({Louisiana(La)}, 190) ({New Orleans}, 179) ({Indiana(In)}, 165)};

            \end{groupplot}
        \end{tikzpicture}
    }
    \vspace{-23pt}
    \caption{Distribution of user profile attributes inferred by \llmrep\ for ML-1M, A-Books, and Yelp.}
    \vspace{-0.35cm}
    \label{fig:RQ1_prof_combined}
\end{figure}
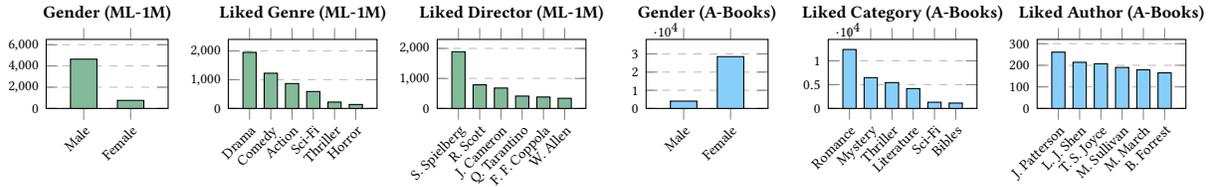
\begin{figure}[t]
\vspace{-0.2cm}
\small
    \centering
    \resizebox{\columnwidth}{!}{ 
        \begin{tikzpicture}
            \begin{groupplot}[
                group style={
                    group size=3 by 1, 
                    horizontal sep=0.6cm, 
                },
                ybar,
                width=0.36\columnwidth, 
                height=2.3cm,
                ylabel style={font=\tiny},
                xtick=data,
                xticklabel style={
                    rotate=45,
                    anchor=north east,
                    font=\fontsize{3.8}{4.8}\selectfont,
                    inner sep=1pt
                },
                yticklabel style={font=\tiny},
                ymin=0,
                enlarge x limits=0.1,
                ymajorgrids=true,
                grid style=dashed,
                title style={font=\fontsize{5.5}{6.5}\selectfont, yshift=-2pt},
            ]

            \nextgroupplot[title={\textbf{Genre (ML-1M)}}, 
                bar width=3pt, 
                symbolic x coords={Drama, Comedy, Romance, Action, Crime, Documentary, Musical, Mystery, Thriller}, 
                ymax=4500]
            \addplot[fill=seagreen!60!white, draw=black] coordinates {
                (Drama, 3899) (Comedy, 1047) (Romance, 385) (Action, 314) 
                (Crime, 290) (Documentary, 174) (Musical, 171) (Mystery, 131) (Thriller, 33)
            };

            \nextgroupplot[title={\textbf{Category (A-Books)}}, 
                bar width=3pt, 
                symbolic x coords={Contemporary, Fiction, Mystery, Thrillers, Fantasy, Literature, Paranormal, Romantic, Women's Fiction, United States}, 
                ymax=1250]
            \addplot[fill=lightskyblue, draw=black] coordinates {
                (Contemporary, 1073) (Fiction, 734) (Mystery, 556) (Thrillers, 373) 
                (Fantasy, 345) (Literature, 305) (Paranormal, 305) 
                (Romantic, 246) (Women's Fiction, 236) (United States, 222)
            };

            \nextgroupplot[title={\textbf{Category (Yelp)}}, 
                bar width=3pt, 
                symbolic x coords={Restaurants, Shopping, Food, Nightlife, Bars, Beauty, Automotive, Medical, Home}, 
                ymax=1500]
            \addplot[fill=orange!40!white, draw=black] coordinates {
                (Restaurants, 1200) (Shopping, 850) (Food, 600) (Nightlife, 450) 
                (Bars, 400) (Beauty, 350) (Automotive, 250) (Medical, 200) (Home, 150)
            };

            \end{groupplot}
        \end{tikzpicture}
    }
    \vspace{-23pt} 
    \caption{Distribution of augmented item attributes generated by \llmaug\ for ML-1M, A-Books, and Yelp.}
    \vspace{-15pt}
    \label{fig:RQ1_augmentation}
\end{figure}
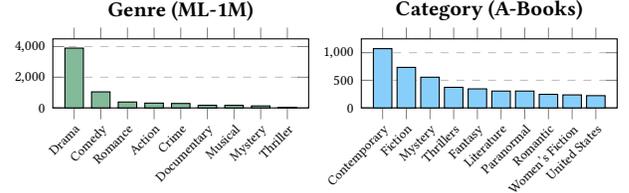

\vspace{1mm}
\noindent\textbf{Implementation Details.}
All baselines are implemented following the protocols reported in their original papers ~\cite{augmentation_Wang_aug_www24_4, llmrec_Wei_comb_aug_prof_rep_wsdm24_12, allmrec_Kim_comb_rec_rep_kdd24_5}.
All training and inference were run on a Linux server with an NVIDIA RTX 6000 Ada GPU.
Table~\ref{Table:llm_statistics} summarizes the LLM backbones and associated generation costs.
Complete implementation details, prompts, data preprocessing pipelines, and hyperparameter configurations are provided in the \href{https://sites.google.com/view/llm4rs-diag/main}{\color{blue}{\ul{online appendix}}}. 

\vspace{1mm}
\noindent\textbf{Feedback-Loop Configuration.}
Following Section~\ref{sec:FeedbackLoop-Pipeline}, we construct temporal feedback-loop splits using a \textbf{cutoff ratio} $t$ and the \textbf{number of periods} $N$.
We choose these configurations to ensure a sufficient number of \textbf{common users} appearing in both the initial training set and subsequent feedback-loop periods. 
For ML-1M, we use $t=0.8$ and $N=5$, resulting in 1,143 common users.
This higher cutoff reflects the smaller user base and shorter activity span per user in ML-1M, which reduces the number of users active across both training and feedback-loop periods.
For A-Books and Yelp, we use $t=0.5$ and $N=5$, yielding 31,445 and 9,102 common users, respectively.
We verified that the overall trends remain consistent under alternative configurations (\eg, different $t$ and $N$ values).

For the \textbf{mixing coefficient}, we evaluate $\alpha \in \{1.0, 0.5\}$.
In the main experiments, we focus on $\alpha = 1.0$ to isolate long-term risk propagation under \textit{fully} LLM-driven feedback loops.
This setting serves as a controlled stress-test environment for diagnosing recursive risk accumulation rather than a realistic deployment condition.
Additional analyses under $\alpha = 0.5$ are deferred to Section~\ref{sec:discuss}. 

\vspace{1mm}
\noindent\textbf{Evaluation Metrics.}
Following Section~\ref{sec:diagnostics}, we evaluate risks from three categories: (i) \textit{bias}, (ii) \textit{hallucination}, and (iii) \textit{long-term effects}.

To diagnose \textbf{LLMGC bias}, we analyze the \textit{distributions} of generated user/item attributes and augmented interactions to identify skewed or dominant patterns. 
When ground-truth attributes are available (\eg, gender and occupation in ML-1M), we additionally compare generated distributions against empirical distributions to assess whether LLMGC preserves, amplifies or introduces new distributional patterns. 
To diagnose \textbf{ranking bias}, we define the \textit{popularity score} of each item as the number of interactions observed in the training set.
For each user $u$, we compare the average popularity of recommended items with that of ground-truth interactions, and define their difference as the \textit{popularity gap}.
We further track the temporal evolution of both LLMGC distributions and popularity gaps across feedback-loop periods.

To diagnose \textbf{FEF}, we define the FEF rate as 
$1-\frac{\lvert \mathcal{O}_{\text{LLM}} \cap \mathcal{O}_{\text{gt}} \rvert}{\lvert \mathcal{O}_{\text{LLM}} \rvert},$ where $\mathcal{O}_{\text{LLM}}$ denotes generated attributes or recommendation items and $\mathcal{O}_{\text{gt}}$ denotes the corresponding ground-truth set.
To diagnose \textbf{LC}, we define the LC rate as 
$\frac{\sum \mathbb{I}\!\left[\mathcal{O}^{(1)}_{\text{LLM}} \neq \mathcal{O}_{\text{LLM}}^{(2)}\right]}{\lvert \mathcal{O}_{\text{gt}} \rvert},$
where $\mathcal{O}^{(1)}_{\text{LLM}}$ and $\mathcal{O}^{(2)}_{\text{LLM}}$ denote two independently generated outputs under identical conditions.
These metrics quantify factual errors and generation instability across repeated LLM outputs, while also tracking their temporal evolution across feedback-loop periods.

Lastly, we analyze the evolution of user and item embeddings across feedback-loop periods to diagnose \textbf{long-term structural separation}.
Using $k$-means clustering on the final-period embeddings, we obtain reference cluster labels and track the same users and items backward across earlier periods. 
We empirically select $k=2$, which yields the highest silhouette score among $k \in \{2, \ldots, 10\}$.
We then measure changes in Euclidean distances between cluster centroids over time. 
To interpret the observed separation, we additionally analyze: (i) preference-level differences between clusters, (ii) popularity-based exposure patterns, and (iii) attention allocation toward previously injected LLM-generated items.
These analyses help determine whether the separation is better explained by conventional preference/popularity effects or by self-reinforcing reliance on LLM-injected signals.

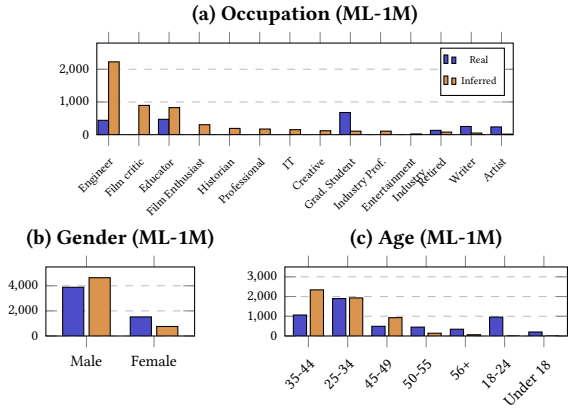
\begin{figure}[t]
\small
    \centering
    \begin{tikzpicture}
        \begin{axis}[
            ybar,
            title={\textbf{(a) Occupation (ML-1M)}},
            title style={font=\small, yshift=-2pt},
            width=0.4\textwidth,  
            height=2.8cm,
            ylabel style={font=\tiny},
            yticklabel style={font=\tiny},
            ymin=0, ymax=2800,
            ymajorgrids=true,
            grid style=dashed,
            symbolic x coords={tech, critic, edu, enthusiast, historian, professional, it, creative, student, ent_prof, ent_ind, retired, writer, artist},
            xtick={tech, critic, edu, enthusiast, historian, professional, it, creative, student, ent_prof, ent_ind, retired, writer, artist},
            xticklabels={
                {Engineer}, {Film critic}, {Educator}, {Film Enthusiast}, {Historian}, {Professional}, {IT}, {Creative}, {Grad. Student}, {Industry Prof.}, {Entertainment\\Industry}, {Retired}, {Writer}, {Artist}
            },
            xticklabel style={rotate=45, anchor=north east, font=\fontsize{4.5}{5.5}\selectfont, align=right},
            enlarge x limits=0.03,
            legend style={
                at={(0.98,0.95)}, 
                anchor=north east, 
                font=\fontsize{4}{5}\selectfont, 
                inner sep=1pt,
                legend image post style={scale=0.5}
            }
        ]
            \addplot[fill=blue!70!black!70, bar width=4pt, bar shift=-2pt] coordinates {(tech, 439) (edu, 469) (student, 677) (retired, 132) (writer, 250) (artist, 236)};
            \addlegendentry{Real}
            \addplot[fill=orange!80!black!70, bar width=4pt, bar shift=2pt] coordinates {(tech, 2226) (critic, 895) (edu, 827) (enthusiast, 304) (historian, 191) (professional, 172) (it, 153) (creative, 120) (student, 107) (ent_prof, 107) (retired, 78) (writer, 49) (ent_ind, 21) (artist, 16)};
            \addlegendentry{Inferred}
        \end{axis}
    \end{tikzpicture}

    \vspace{-0.2cm} 

    \begin{tikzpicture}
        \begin{groupplot}[
            group style={
                group size=2 by 1,
                horizontal sep=1.2cm, 
            },
            ybar,
            height=2.5cm,
            ylabel style={font=\tiny},
            yticklabel style={font=\tiny},
            ymin=0,
            ymajorgrids=true,
            grid style=dashed,
            title style={font=\small, yshift=-2pt},
        ]

        \nextgroupplot[
            title={\textbf{(b) Gender (ML-1M)}},
            width=0.2\textwidth,
            symbolic x coords={Male, Female},
            xtick={Male, Female},
            ymax=5500,
            bar width=8pt,
            enlarge x limits=0.6,
            xticklabel style={font=\fontsize{6}{7}\selectfont}
        ]
            \addplot[fill=blue!70!black!70, bar shift=-5pt] coordinates {(Male, 3880) (Female, 1520)};
            \addplot[fill=orange!80!black!70, bar shift=5pt] coordinates {(Male, 4643) (Female, 757)};

        \nextgroupplot[
            title={\textbf{(c) Age (ML-1M)}},
            width=0.3\textwidth,
            symbolic x coords={35-44, 25-34, 45-49, 50-55, 56+, 18-24, Under 18},
            xtick={25-34, 35-44, 45-49, 18-24, 50-55, 56+, Under 18},
            ymax=3500,
            bar width=5pt,
            enlarge x limits=0.1,
            xticklabel style={rotate=45, anchor=north east, font=\fontsize{6}{7}\selectfont}
        ]
            \addplot[fill=blue!70!black!70, bar shift=-3.2pt] coordinates {(35-44, 1063) (25-34, 1896) (45-49, 490) (50-55, 450) (56+, 343) (18-24, 956) (Under 18, 202)};
            \addplot[fill=orange!80!black!70, bar shift=3.2pt] coordinates {(35-44, 2339) (25-34, 1930) (45-49, 928) (50-55, 140) (56+, 63) (18-24, 0) (Under 18, 0)};

        \end{groupplot}
    \end{tikzpicture}

    \vspace{-0.4cm}
    \caption{Ground-truth attributes vs. attributes inferred by \llmrep\ for ML-1M: `\textit{Real}' denotes ground-truth labels, while `\textit{Inferred}' indicates LLM-inferred attributes.}
    \vspace{-0.4cm}
    \label{fig:RQ1_profile_gt}
\end{figure}



\begin{table}[t] 
\small
\centering
\caption{Analysis of FEF rates in user profiles generated by \llmrep\ on ML-1M. Note that ground-truth labels are available only for occupation, age, and gender. 
}
\vspace{-0.2cm}
\label{tab:RQ1_factual_errors}
\resizebox{0.72\columnwidth}{!}{
    \begin{tabular}{@{}cccc@{}}
    \toprule
    \textbf{Dataset} & \textbf{\# Target Users} & \textbf{Attribute} & \textbf{FEF Rate (\%)} \\ \midrule
    \multirow{3}{*}{\textbf{ML-1M}} & \multirow{3}{*}{5,400} & Occupation & 93.16\% \\
      &  & Age & 73.68\% \\
      &  & Gender & 25.35\% \\ \bottomrule
    \end{tabular}
}
\vspace{-0.35cm}
\end{table}

\subsection{Experimental Results}
We report representative findings here, while complete results and figures are provided in the \href{https://sites.google.com/view/llm4rs-diag/main}{\color{blue}{\ul{online appendix}}}.

\begin{figure*}[t]
\centering
\includegraphics[width=1\linewidth]{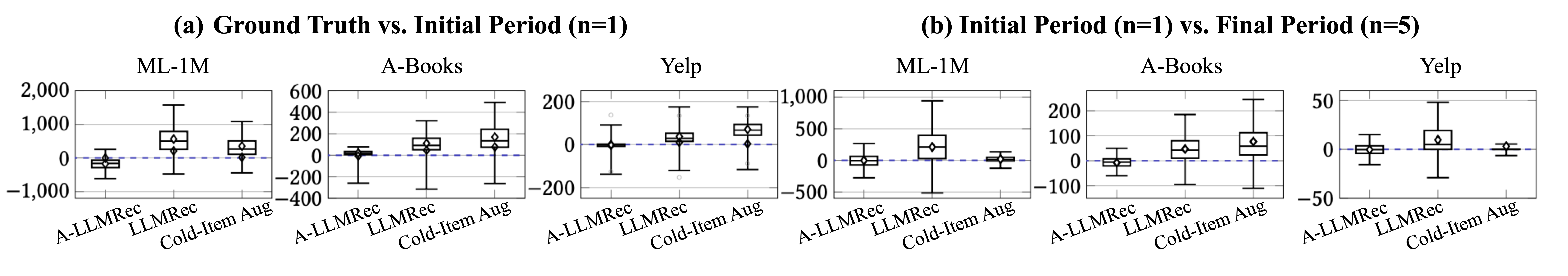}
\vspace{-25pt}
\caption{Distribution of the popularity gap across all roles: 
    \textbf{(a)} popularity gap between ground truth interactions ($\mathcal{D}^{\text{gt}}$) and ranked lists at the initial period ($n=1$), and \textbf{(b)} popularity gap between ranked lists from the initial ($n=1$) and final ($n=5$) periods.
    Positive (negative) values indicate increased (decreased) exposure to popular items relative to the ground truth. 
    }
\label{fig:RQ3_popularity_6by1}
\vspace{-0.4cm}
\end{figure*}

\subsubsection{Phase 1 Diagnosis (RQ1)}
\hfill

\vspace{1mm}
\noindent\textbf{Bias in LLMGC.}
Figure~\ref{fig:RQ1_prof_combined} shows the attribute distributions of user profiles generated by LLMRec, while Figure~\ref{fig:RQ1_augmentation} reports the genre/category distributions of augmented interactions produced by Cold-Item Aug. 
Across both settings, the generated distributions are \textit{highly skewed}, with a small subset of inferred attributes, genres, or categories dominating the generated outputs.

We further compare generated user attributes against ground-truth distributions available in ML-1M (\ie, gender, age, and occupation).
As shown in Figure~\ref{fig:RQ1_profile_gt}, two representative patterns emerge.
First, \textit{bias amplification} occurs when LLMGC follows existing trends while increasing their imbalance, such as inflated dominant gender groups (Figure~\ref{fig:RQ1_profile_gt}-(b)). 
Second, \textit{distributional shift} arises when LLMGC introduces new systematic patterns absent from the original data, such as the over-representation of specific occupations that do not reflect the empirical population (Figures~\ref{fig:RQ1_profile_gt}-(a) and -(c)).

\begin{table}[t] 
\small
\centering
\caption{Analysis of LC rates in user profiles generated by \llmrep\ on ML-1M, A-Books, and Yelp}
\vspace{-0.2cm}
\label{tab:RQ1_logical_contradiction}
\resizebox{0.84\columnwidth}{!}{
    \begin{tabular}{@{}ccccc@{}}
    \toprule
    \textbf{Dataset} & \textbf{\# Target Users} & \textbf{Attribute} & \textbf{LC Rate (\%)} \\ \midrule
    \multirow{8}{*}{\textbf{(a) ML-1M}} &  \multirow{8}{*}{5,400} & Age & 21.09\% \\
       &  & Liked Director & 20.92\% \\
       &  & Occupation & 15.38\% \\
       &  & Disliked Genre & 14.27\% \\
       &  & Country & 8.59\% \\
       &  & Liked Genre & 4\% \\
       &  & Gender & 1.68\% \\ \midrule
    \multirow{6}{*}{\textbf{(b) A-Books}} &  \multirow{6}{*}{31,445} & Age & 15.05\% \\
      &  & Disliked category & 14.71\% \\
      &  & Liked Author & 3.81\% \\
      &  & Liked Category & 3.31\% \\
      &  & Gender & 0.67\% \\
      &  & Country & 0.14\% \\ \midrule
    \multirow{6}{*}{\textbf{(c) Yelp}} &  \multirow{6}{*}{12,025} & Disliked category & 17.72\% \\
      &  & Occupation & 14.46\% \\
      &  & Age & 10.83\% \\
      &  & Liked category& 8.64\% \\
      &  & Gender & 3.25\% \\
      &  & Preferred city & 0.97\% \\ \bottomrule
      
    \end{tabular}
}
\vspace{-0.35cm}
\end{table}

\vspace{1mm}
\noindent\textbf{Hallucination in LLMGC.}
Table~\ref{tab:RQ1_factual_errors} reports substantial FEF rates across generated user attributes in ML-1M from 25.35\% (Gender) to 93.16\% (Occupation).
A representative case is shown in Figure~\ref{fig:RQ1_profile_gt}-(a), where many users are inferred as `\textit{film critic},' an occupation absent from the dataset.
This pattern suggests that LLMGC can fabricate plausible but non-existent attributes influenced by domain-specific narratives.
We further evaluate LC rates across ML-1M, A-Books, and Yelp.
As shown in Table~\ref{tab:RQ1_logical_contradiction}, LC rates remain consistently non-trivial across datasets and attributes, indicating that LLMGC is not only prone to fabrication but also sensitive to stochastic variation under identical inputs.

\subsubsection{Phase 2 Diagnosis (RQ2)}
\hfill

\vspace{1mm}
\noindent\textbf{Bias in Ranking.}
Figure~\ref{fig:RQ3_popularity_6by1}-(a) shows that the popularity gap between ground-truth interactions and ranked lists varies substantially across LLM roles. 
LLMRec and Cold-Item Aug consistently increase popularity bias across all three datasets, indicating that LLMGC can reinforce both pre-existing skews and newly introduced biases observed in Phase 1.
In contrast, A-LLMRec maintains or slightly reduces the popularity gap, potentially due to hallucinated recommendations that increase apparent diversity.

\begin{table}[t]
\centering
\caption{FEF analysis in ranked lists produced by \llmrec: 
`\textit{\# Recommended Items}' denotes the total number of recommended items over common users,
and 
`\textit{FEF Case}' reports examples of hallucinated items.}
\vspace{-0.2cm}
\label{tab:RQ2_a_llmrec_fef}
\resizebox{0.98\columnwidth}{!}{
    \begin{tabular}{@{}cccc@{}}
    \toprule
    \textbf{Dataset} & \textbf{\# Recommended Items} &  \textbf{FEF Rate (\%)} & \textbf{FEF Case}\\ \midrule
    \textbf{(a) ML-1M} & 104,540 &  4.07\% & \begin{tabular}[c]{@{}c@{}}First of all (2000), \\ Golden Boy, The (1986)\end{tabular} \\ \midrule
    \textbf{(b) A-Books} &  440,917 & 7.40\% & \begin{tabular}[c]{@{}c@{}}Hyperion (Hyperspace), \\ King and Maxwell (Wingman)\end{tabular} \\ \midrule
    \textbf{(c) Yelp} &  52,027 & 13.14\% & \begin{tabular}[c]{@{}c@{}}BistroChili's, \\ Buckley's Debut\end{tabular} \\ \bottomrule
    \end{tabular}
}
\vspace{-0.2cm}
\end{table}

\vspace{1mm}
\noindent\textbf{Hallucination in Ranking.}
Note that hallucinated recommendation items are evaluated only for A-LLMRec, since LLMRec and Cold-Item Aug use LLMs only during content generation rather than direct recommendation generation.
Table~\ref{tab:RQ2_a_llmrec_fef} reports non-trivial FEF rates in ranked lists generated by A-LLMRec, ranging from 4.07\% to 13.14\%.
Combined with the popularity analysis in Figure~\ref{fig:RQ3_popularity_6by1}, these results suggest a potential trade-off in open-ended \llmrs: while direct LLM-based recommendation generation may reduce popularity concentration, it can simultaneously introduce unreliable or non-existent recommendation items. 

\begin{figure*}[t]
\centering
\includegraphics[width=1\linewidth]{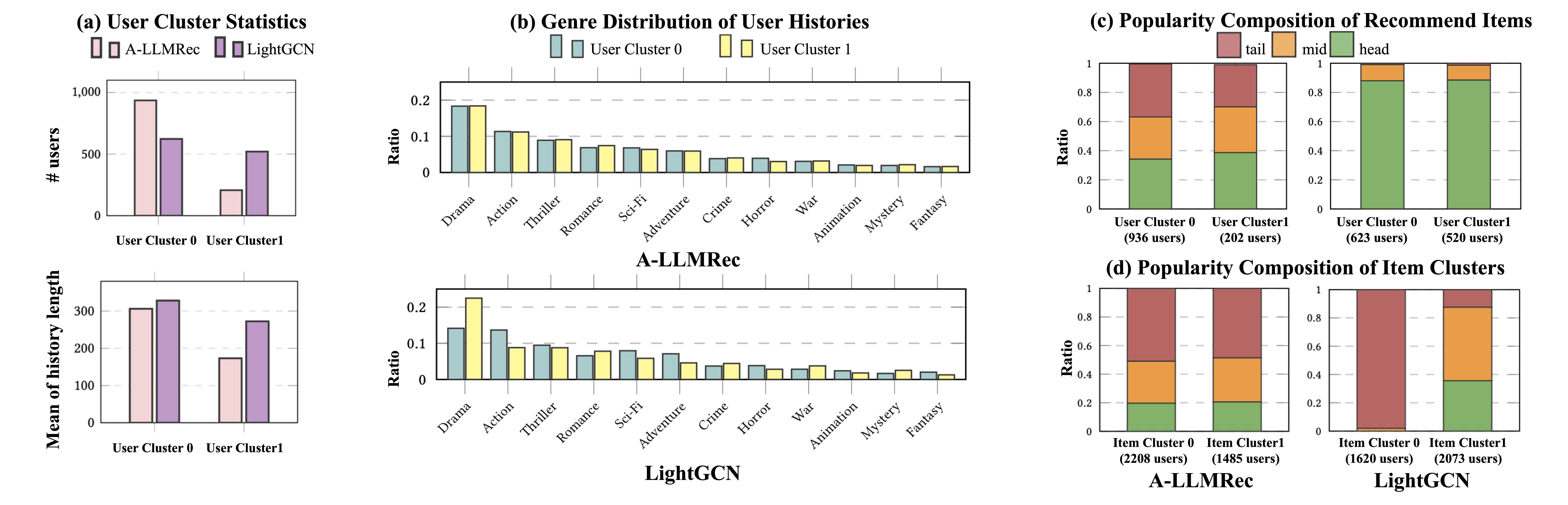}
\vspace{-25pt}
\caption{
Cluster analysis of user and item embeddings learned by A-LLMRec and LightGCN at the final feedback period on ML-1M:
(a) user cluster size and interaction history length,
(b) genre distribution of user interaction histories,
(c) head/mid/tail composition of recommended items per user cluster, 
and 
(d) head/mid/tail composition of item clusters.
}
\label{fig:RQ4_cluster_analysis}
\vspace{-0.4cm}
\end{figure*}

\subsubsection{Phase 3 Diagnosis (RQ3 \& RQ4)}
\hfill

\vspace{1mm}
\noindent\textbf{Accumulation of Bias and Hallucination.}
Across all \llmrs, the bias and hallucination signals identified in earlier phases generally persist throughout feedback-loop iterations, although their trajectories vary across datasets. 
For example, the gender imbalance observed in Figure~\ref{fig:RQ1_prof_combined} becomes further amplified in ML-1M (male: 85.90\% $\rightarrow$ 86.80\%) and A-Books (male: 12.40\% $\rightarrow$ 11.90\%), while Yelp exhibits relatively weaker amplification trends. 
Similarly, most generated attribute distributions preserve their dominant patterns over time.
As shown in Figure~\ref{fig:RQ3_popularity_6by1}-(b), the popularity gap in aggregated recommendation outputs continues to increase across feedback periods.
Hallucination signals also remain consistently high over time.
For example, on ML-1M, the FEF rate for Occupation remains above 93\%, while the LC rate for Age remains around 21\% across feedback periods.
These results suggest that risks introduced during earlier stages are not naturally corrected through recursive feedback, but can instead persist and accumulate over time.

\begin{figure}[t] 
\centering
\includegraphics[width=\columnwidth]{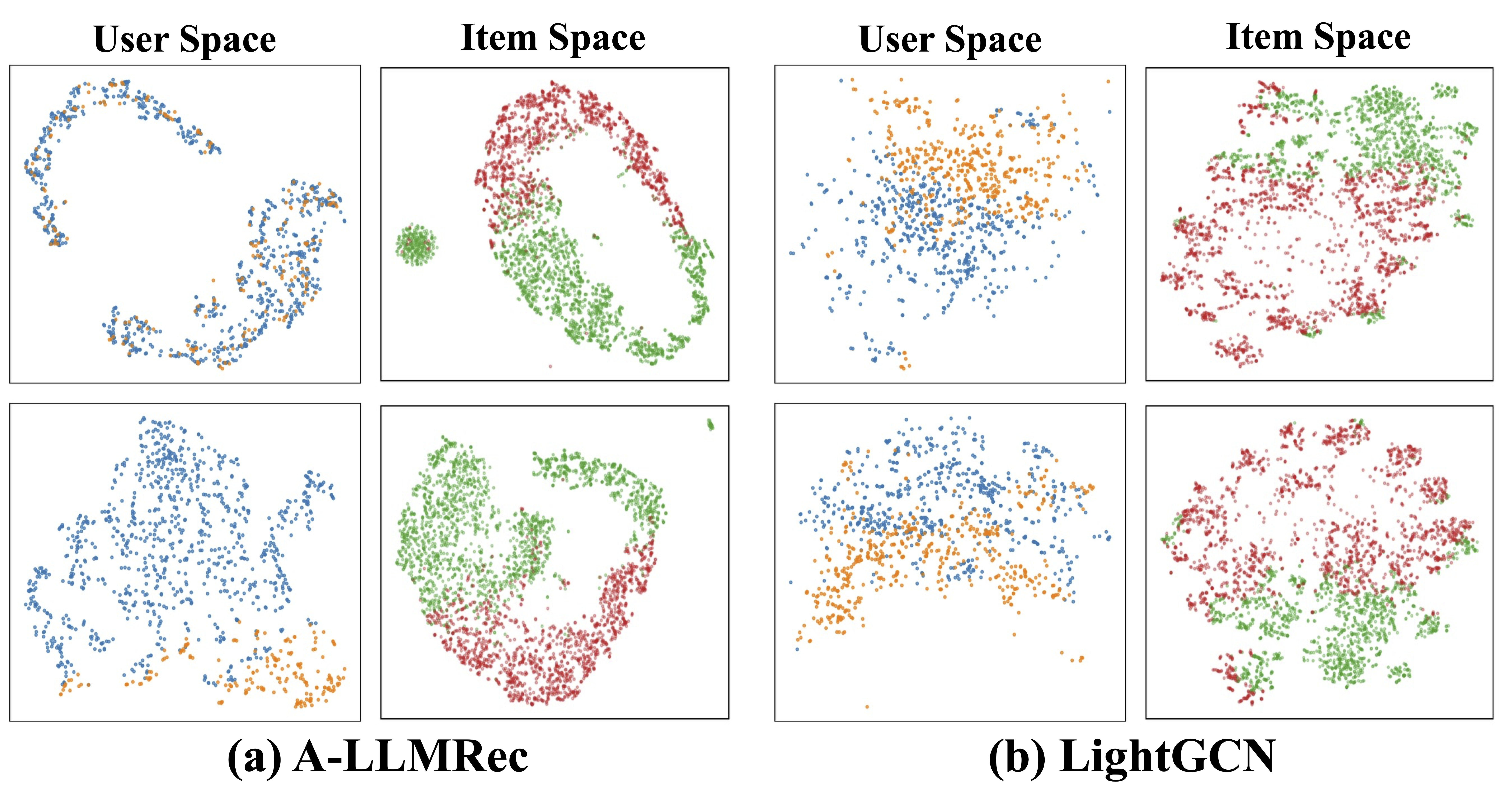} 
\\ \vspace{-0.3cm} 

\caption{t-SNE visualization of user and item embeddings under the feedback loop.
The top row shows the  initial period ($n=1$), and the bottom row shows the final feedback period ($n=5$).
Points are colored according to cluster assignments obtained by $k$-means clustering performed at $n=5$ and propagated to earlier stages for the same users and items.}

\label{fig:RQ4_pol_ml1m}
\vspace{-0.4cm}
\end{figure}

\vspace{1mm}
\noindent\textbf{Long-Term Structural Separation.}
Following Section~\ref{sec:setting}, we apply $k$-means clustering ($k=2$) to the final-period user and item embeddings (\ie, $n=5$), and track the same groups backward across earlier feedback periods.
Figure~\ref{fig:RQ4_pol_ml1m}-(a) visualizes the embedding evolution of A-LLMRec on ML-1M from the initial stage ($n=1$) to the final feedback period ($n=5$).
As feedback iterations progress, the learned user and item groups become increasingly separated in the embedding space, particularly in the user representations.
To quantify this effect, we measure Euclidean distances between group centroids over time.
For A-LLMRec on ML-1M, the user-group distance increases from 3.73 to 9.29, while similar trends are consistently observed across other \llmrs. 

To assess whether this behavior is specific to \llmrs, we apply the same feedback-loop pipeline to a conventional collaborative filtering (CF)-based RS, LightGCN~\cite{lightgcn}.
As shown in Figure~\ref{fig:RQ4_pol_ml1m}-(b), LightGCN exhibits substantially weaker or no systematic separation trends under the same setting.
These observations suggest that recursive feedback in \llmrs\ may induce  structural effects that differ qualitatively from those in conventional CF-based RS.

\vspace{1mm}
\noindent\textbf{Separation Beyond Human-Driven Patterns.}
While Figure~\ref{fig:RQ4_pol_ml1m} confirms growing structural separation in \llmrs, centroid distances alone do not explain its underlying mechanism. 
Prior studies on feedback loops in conventional CF-based RS have shown that repeated retraining can reinforce popularity-oriented exposure and progressively skew user interaction patterns~\cite{cf_feedback_loop1, cf_feedback_loop2}. 
As a result, structural separation in CF-based RS is often associated with human-driven factors such as preference divergence or popularity concentration.
We therefore examine whether the separation observed in \llmrs\ follows these conventional patterns.

We first analyze the activity characteristics of the resulting user clusters.
As shown in Figure~\ref{fig:RQ4_cluster_analysis}-(a), LightGCN produces relatively balanced user groups with comparable interaction histories.
In contrast, A-LLMRec produces highly imbalanced user clusters despite showing comparable interaction history lengths across groups, suggesting that the observed separation is not primarily driven by differences in user activity.
Therefore, we further examine whether the user clusters exhibit different preference patterns (Figure~\ref{fig:RQ4_cluster_analysis}-(b)). 
Under LightGCN, the two user clusters show clearly distinguishable genre preferences based on users' historical interactions.
However, A-LLMRec exhibits nearly identical genre distributions across clusters, indicating weak preference-level differentiation. 

We further analyze the popularity composition of items recommended to each user cluster, where items are partitioned into head, mid, and tail groups according to their interaction frequencies in the training data.
As shown in Figure~\ref{fig:RQ4_cluster_analysis}-(c), LightGCN exhibits clear differences in head/mid/tail exposure across clusters, reflecting the popularity-oriented behavior of CF-based RS.
In contrast, A-LLMRec shows highly similar popularity compositions across user clusters.
A similar pattern is observed in the item clusters (Figure~\ref{fig:RQ4_cluster_analysis}-(d)). 
While LightGCN produces item clusters with clearly different head/mid/tail compositions, A-LLMRec produces item clusters with substantially similar popularity compositions.

\begin{figure}[t]
    \centering
    \begin{tikzpicture}
        \begin{axis}[
            width=6cm, height=2.5cm,
            xlabel={Period},
            ylabel={Ratio (\%)},
            ylabel style={
                font=\fontsize{8}{9}\selectfont,
                align=center 
            },
            xlabel style={font=\fontsize{8}{9}\selectfont},
            xmin=0.85, xmax=4.15,
            ymin=70, ymax=100,
            xtick={1,2,3,4},
            ytick={70,80,90,100},
            grid=both,
            grid style={dashed, gray!40, line width=0.4pt},
            axis line style={line width=1.0pt},
            tick style={line width=1.0pt},
        ]
        \addplot[
            color=blue, mark=o,
            mark options={scale=0.8, line width=0.6pt},
            line width=1.0pt, dashed,
        ] coordinates {
            (1,86) (2,92) (3,90) (4,87)
        };
        \end{axis}
    \end{tikzpicture}
    \vspace{-10pt}
    \caption{Fraction of users whose top-attended history item is an injected LLM-generated item across feedback periods.}
    \vspace{-10pt}
\label{fig:RQ4_attention}
\end{figure}

\vspace{1mm}
\noindent\textbf{Self-Reinforcement of LLM-Injected Signals.}
Since the observed separation in A-LLMRec is not well explained by conventional human-driven patterns, we next examine whether it instead reflects recursive reliance on previously injected LLM-induced signals through attention-based analysis.
Attention analysis is widely used to interpret which contextual evidence is most emphasized during LLM generation behavior~\cite{attention_ref}.
In most \llmrs, including A-LLMRec, the user's interaction history is provided as part of the prompt context during recommendation generation, where previously injected LLM-generated items may become mixed with the user's actually consumed items across feedback periods.

To analyze whether A-LLMRec relies on its own injected signals, we identify the history item receiving the highest attention score during inference, and measure the fraction of users whose top-attended item corresponds to a previously injected LLM-generated item.
As shown in Figure~\ref{fig:RQ4_attention}, this ratio remains consistently high across feedback periods (86\%--92\%). 
These results suggest that previously injected LLM-generated items can become dominant contextual evidence during subsequent recommendation generation, potentially creating self-reinforcing feedback dynamics.

\vspace{1mm}
In summary, the long-term structural separation observed in \llmrs\ appears to differ qualitatively from the cluster separation typically observed in conventional CF-based RS, and instead appears more closely associated with recursive reliance on LLM-generated signals.
Since our goal is to diagnose and verify the existence of potentially distinct structural effects under recursive LLM-driven feedback loops, a deeper understanding of how such dynamics emerge remains an important direction for future work.

\section{Extended Analyses} \label{sec:discuss} 
Full results are provided in the \href{https://sites.google.com/view/llm4rs-diag/main}{\color{blue}{\ul{online appendix}}}.

\vspace{1mm}
\noindent\textbf{Performance Degradation under Recursive Feedback.} 
Although our primary goal is risk diagnosis rather than recommendation optimization, we additionally examine how recommendation performance evolves under recursive feedback loops.
Under the default setting ($\alpha=1$), where all injected interactions follow LLM recommendations, \llmrs\ generally exhibit gradual performance degradation over feedback periods.
For example, on ML-1M, A-LLMRec degrades Recall by up to $6.19\%$ from $n=1$ to $n=5$.
These results suggest that recursive exposure to LLM-generated interactions can gradually distort the behavioral signals used for subsequent recommendation learning.

\vspace{1mm}
\noindent\textbf{Preserving Human-Driven Interactions in Feedback Loops.}
We next examine whether retaining authentic human-driven interactions can mitigate long-term risks under recursive feedback loops.
To this end, we compare $\alpha=1$ with $\alpha=0.5$, where half of the injected interactions are replaced by actually consumed items.
Under $\alpha=0.5$, the bias and hallucination patterns observed under $\alpha=1$ remain observable, although generally at weaker levels, suggesting that LLM-generated signals continue to influence the feedback loop even when partially mixed with authentic user interactions.

However, we observe different performance dynamics under recursive feedback.
On ML-1M, A-LLMRec improves from $n=1$ to $n=5$ under $\alpha=0.5$ (Recall $+23.74\%$, Hit $+4.79\%$), in contrast to the degradation observed under $\alpha=1$.
Consistently, the attention-based analysis shows that the fraction of users whose top-attended history item corresponds to a previously injected LLM-generated item decreases from 86--92\% under $\alpha=1$ to 45--65\% under $\alpha=0.5$. 
These findings suggest that retaining authentic user-consumed items can partially anchor the feedback loop to human behavioral signals, reducing the dominance of LLM-generated interactions in subsequent recommendation learning. 

\vspace{1mm}
\noindent\textbf{Risk Consistency across LLM Backbones.}
We additionally examine whether the diagnosed risks persist across different LLM backbones by replacing the default backbone with \texttt{Claude-Haiku-4-5} and \texttt{Qwen3-14B} in the LLM content generation phase, which serves as the primary source of downstream feedback signals. 
As a result, the overall patterns of bias and hallucination remain consistently observable. 
For example, on ML-1M, \texttt{Claude-Haiku-4-5} generates a substantially more imbalanced gender distribution than \texttt{Qwen3-} \texttt{14B}.
Moreover, both models produce occupation labels outside the ML-1M taxonomy, including ``blue-collar worker'' and ``film enthusiast''.
These observations suggest that the risks identified by the proposed framework are not artifacts of a specific LLM backbone, but can recur across different models in qualitatively similar forms.

\section{Conclusion} \label{sec:concl} In this work, we presented a functional taxonomy of \llmrs\ and formulated their risks from a recursive feedback-loop perspective.
Building on this formulation, we proposed a role-aware, three-phase diagnostic framework with a controlled experimental pipeline. 
Our longitudinal analyses show that LLM-generated signals can propagate bias and hallucination across feedback iterations, amplifying popularity bias and inducing representation drift and structural separation over time.
These findings highlight the need for feedback-aware evaluation beyond short-term recommendation accuracy.

Future work can incorporate additional LLM-specific risks (\eg, prompt sensitivity and sycophancy), evaluate mitigation strategies, and extend the framework to other iterative LLM-powered systems. 
We hope this work provides a useful foundation for diagnosing long-term risks in recursive LLM-driven environments.

\begin{acks}
This work was supported by the Institute of Information \& Communications Technology Planning \& Evaluation (IITP) grants funded by the Korea government (MSIT) under the AI Star Fellowship Program (UNIST) (RS-2025-25442824), the Artificial Intelligence Graduate School Program (UNIST) (No. RS-2020-II201336), the Leading Generative AI Human Resources Development Program (IITP-2026-RS-2024-00360227), the Innovative Human Resource Development for Local Intellectualization Program (IITP-2026-RS-2022-00156361), and the InnoCORE program of the Ministry of Science and ICT (UNIST 1.260006.01). 
This work was also supported in part by NSF awards \#2555559 and \#2618891.
\end{acks}

\clearpage
\section*{GenAI Usage Disclosure} \label{sec:genai} We used generative AI tools (e.g., ChatGPT) solely for linguistic improvements, including grammar correction and sentence refinement. In addition, AI tools were used to assist code development by generating code snippets and suggesting debugging solutions. Importantly, generative AI tools were not used for idea generation, experimental design, data analysis, or result interpretation.

\balance
\bibliographystyle{ACM-Reference-Format}
\bibliography{bibliography}

\clearpage


\end{document}